\documentclass{elsart}

\usepackage{amssymb}
\usepackage[square,comma]{natbib}
\usepackage{graphicx}

\begin{document}
\begin{frontmatter}



\def\qed{\'\'$\Box$}

\title{Direct definition of a ternary infinite square-free sequence}


\author{Tetsuo Kurosaki}
\ead{kurosaki@monet.phys.s.u-tokyo.ac.jp}

\address{Department of Physics, Graduate School of Science, University of Tokyo, Hongo, Bunkyo-ku, Tokyo 113-0033, Japan}
\address{Bank of Japan, Hongoku-cho, Nihonbashi, Chuo-ku, Tokyo 103-8660, Japan}

\begin{abstract}
We propose a new ternary infinite (even full-infinite) square-free sequence. The sequence is defined both by an iterative method and by a direct definition. Both definitions are analogous to those of the Thue-Morse sequence. The direct definition is given by a deterministic finite automaton with output. In short, the sequence is automatic. 
\end{abstract}


\end{frontmatter}


\section{Introduction}
\label{}

First of all, we introduce several notations to be used in this paper. Let $\Sigma$ be a non-empty and finite set called \textit{alphabet} and $\Sigma^*$ be the free monoid generated by $\Sigma$. A sequence consists of elements of the free monoid, $i$-th element of which is usually assigned a symbol with a subscript $i$. Specifically, a sequence $\mathbf{a}$ over $\Sigma$, or $\mathbf{a} \in \Sigma^*$, with the length $n+1$ is expressed as $\mathbf{a}=\{a_i\}_{i=0}^n \equiv a_0a_1\cdots a_n$ where $\forall a_i\in \Sigma$. We usually make the numbering of elements of a sequence start with 0, whereas this numbering will be altered later. A sequence can be either empty or infinite. Concatenation of two finite sequences is denoted by juxtaposition. For example, $\{a_i\}_{i=0}^n \{a_{n+i+1}\}_{i=0}^m=\{a_i\}_{i=0}^{n+m+1}$. We say that a sequence $\mathbf{a}$ contains another sequence $\mathbf{u}$, called a factor of $\mathbf{a}$, if there are $\mathbf{r}$ and $\mathbf{r}^*$ such that $\mathbf{a}=\mathbf{r} \mathbf{u} \mathbf{r}^*$. The concept of the factor is defined to an infinite sequence as well.

Thue introduced a simple and mysterious binary infinite sequence $\mathbf{t}=\{t_i\}_{i=0}^\infty \in \{0,1\}^*$ and discussed the combinatorial properties of it \cite{thue1,thue2}. It starts such that
\begin{equation}
\mathbf{t}=0110100110010110\cdots . 
\end{equation}
Morse rediscovered the sequence \textbf{t} in the context of differential geometry \cite{morse1, morse2}. After the two pioneers, the sequence \textbf{t} is now called the \textit{Thue-Morse sequence}, which has a long history of research \cite{automatic, allouche, remark, combinatorics}. 

There exist several ways of the mathematical definition of the Thue-Morse sequence $\mathbf{t}$. A popular one is based on a recursive construction. Define an operator $\psi: \{0,1\}^*\to \{0,1\}^*$ which doubles the length of the sequence such that $\psi(\textbf{a})=\mathbf{a} \bar{\mathbf{a}}$, where $\bar{\mathbf{a}}$ is the sequence obtained from $\textbf{a}$ by exchanging 0 and 1. We find that $\psi(0)=01$, $\psi^2(0)=0110$ and $\psi^\infty(0)=\lim_{n\to \infty} \psi^n(0)$ converges toward $\mathbf{t}$ for the ordinary topology, which is a fixed point of $\psi$. Another definition is more direct. Let $s_2(i)$ be the sum of digits in the binary representation of an integer $i$ and then $s_2(i)$ corresponds to $t_i$ under modulo 2. From the latter definition, we can construct a deterministic finite automaton with output (DFAO) which takes as input the binary representation of $i$ and yields $t_i$. This means that the Thue-Morse sequence is 2-automatic \cite{automatic, Cobham}.     

The Thue-Morse sequence is known to appear in various distinct fields such as combinatorics, number theory, group theory, real analysis, information science, solid state physics and so on (see, for instance, a review \cite{allouche}). In terms of combinatorial aspects, a remarkable fact is that a \textit{square-free} sequence can be derived by utilizing the Thue-Morse sequence as the auxiliary sequence \cite{thue1,thue2}. A sequence $\mathbf{u}$ over $\Sigma$ is said to be square-free if $\mathbf{u}$ contains no square, i.e., no two consecutive repeating blocks $\mathbf{w}\mathbf{w}$ with $\mathbf{w}$ a non-empty and finite sequence of $\Sigma^*$. For $i\ge 0$, let $v_i$ be the number of 1's between $i$-th and $(i+1)$-st appearance of 0 in the sequence $\mathbf{t}$. Then, $\mathbf{v}=\{v_i\}_{i=0}^\infty=21020121012\cdots$ is a ternary infinite square-free sequence over $\Sigma=\{0,1,2\}$. The concept of square-free is significant in that it leads to a branch of combinatorics, called \textit{combinatorics on words} \cite{combinatorics}. Many sequences other than $\mathbf{v}$ are known as ternary infinite square-free sequences. In particular, the sequences of Arshon \cite{Arshon}, Leech \cite{Leech}, Zech \cite{Zech} and so on are not only square-free but also automatic, due to the fact that they can be defined by a uniform tag sequence \cite{Cobham}. 

In this paper, we present a new method to define a ternary infinite square-free sequence. First, the sequence is defined by an iterative method and then redefined by a direct definition. Since the direct definition is brought by an automaton operation, we find that the sequence is automatic. Our definitions are quite straightforward in contrast with those of the Thue-Morse sequence.

\section{Construction of a ternary infinite square-free sequence}
\label{Const}

Ternary sequences considered in this section are those over $\{1,2,3 \}$.

Let $\sigma$ and $\rho$ be morphisms or permutations on $\{1,2,3 \}^*$, namely, for $\textbf{a} \in \{1,2,3 \}^*$. They are defined as follows, respectively:

\begin{itemize}
	\item $\sigma(\textbf{a})$ is the sequence obtained from \textbf{a} by exchanging 1 and 2.
	\item $\rho(\textbf{a})$ is the sequence obtained from \textbf{a} by exchanging 2 and 3.
\end{itemize}

In addition, define an operator $\varphi: \{1,2,3\}^*\to \{1,2,3\}^*$ which triples the length of the sequence such that $\varphi(\textbf{a})=\sigma(\textbf{a})\ \textbf{a}\ \rho(\textbf{a})$.

\begin{thm}
\label{square-free}
The sequence $\varphi^n(2)$ is square-free for arbitrary integer $n$.
\end{thm}

It is easy to write explicitly $\varphi^n(2)$ for $n=1,2,3$. They are given by 
\begin{eqnarray*}
\varphi^1(2) &=& 123, \\
\varphi^2(2) &=& 213123132, \\
\varphi^3(2) &=& 123213231213123132312132123, \\
\end{eqnarray*}
and we can see that they are square-free.

The length of $\varphi^n(2)$ is $3^n$. We shall prove two lemmas about $\varphi^n(2) \equiv \{a_i^n\}_{i=0}^{3^n-1}$. These lemmas are very simple, but crucial to prove Theorem \ref{square-free}. The main idea is to decompose $\varphi^n(2)$ into $3^{n-1}$ consecutive ternarys $a_{3k}^na_{3k+1}^na_{3k+2}^n$.

\begin{lem}
\label{ternary}
Each ternary $a_{3k}^na_{3k+1}^na_{3k+2}^n \ (k=0,1,\cdots,3^{n-1}-1)$ is a certain permutation of 123.
\end{lem}

\begin{pf}
Obviously, the iterative operations for the sequence 2 by $\sigma$ and $\rho$, in short, $\varphi$, preserve the property stated by Lemma \ref{ternary}. \qed
\end{pf}

Remark that Lemma \ref{ternary} guarantees that, if two elements of the ternary are provided, one can guess the remaining one element of it.

Define an operator $f$ which extracts $a_{3k+1}^n$ from each ternary $a_{3k}^na_{3k+1}^na_{3k+2}^n$ for $k=0,1,\cdots,3^{n-1}-1$, i.e., $f(\varphi^n(2))=\{a_{3i+1}^n\}_{i=0}^{3^{n-1}-1}$. The idea of extracting the central element of the ternary comes from the analogy of the Cantor set, well known as the fractal set. 

\begin{lem}
\label{central}
The sequence $f(\varphi^n(2))=\{a_{3i+1}^n\}_{i=0}^{3^{n-1}-1}$ is equivalent to $\varphi^{n-1}(2)=\{a_i^{n-1}\}_{i=0}^{3^{n-1}-1}$.
\end{lem}

\begin{pf}
Obviously, $f(\varphi(2))=f(123)=2=\varphi^0(2)$. Assume that, for $n\ge 1$, $f(\varphi^n(2))=\varphi^{n-1}(2)$ is also true. Note that the operators $\varphi$ and $f$ are commutable, and then $f(\varphi^{n+1}(2))=\varphi(f(\varphi^n(2))=\varphi^n(2)$. Therefore, $f(\varphi^n(2))=\varphi^{n-1}(2)$ is proved for arbitrary $n$, by a mathematical induction. \qed
\end{pf}

By using Lemmas \ref{ternary} and \ref{central}, we can prove Theorem \ref{square-free}.

\def\Elproofname{PROOF of Theorem \ref{square-free}.}
\begin{pf}
It is sufficient to prove that $\varphi^n(2)=\{a_i^n\}_{i=0}^{3^n-1}$ does not contain a square with any length $2\ell$, i.e., $\{ a_i^n\}_{i=p}^{p+2\ell-1}$ such that $a_k^n=a_{k+\ell}^n \ (k=p,p+1,\cdots,p+\ell-1)$, for any $p$ and $\ell$. To complete the proof, we divide the problem into several cases, according to the value of $\ell$, whereas we employ a mathematical induction and a reduction to absurdity. It is clear that $\varphi^1(2)=123$ is square-free. Hereafter, we consider $\varphi^{n+1} (2)$ for individual cases divided as to $\ell$, provided that $\varphi^n (2)$ is square-free for $n\ge 1$. 

\underline{\textit{Case 1}:\ \ $\ell=1,2$}

It is obvious that both $\sigma(\varphi^n (2))$ and $\rho(\varphi^n (2))$ are also square-free. Therefore, if $\varphi^{n+1} (2)$ contains a square, it should straddle over the joint between $\sigma(\varphi^n (2))$ and $\varphi^n (2)$, or between $\varphi^n (2)$ and $\rho(\varphi^n (2))$. However, noting that $\varphi^n (2)$ starts and ends such that 123$\cdots$ 123 for odd $n$, and 213$\cdots$132 for even $n$, $\varphi^{n+1} (2)$ is in the form of 213$\cdots$ 213123$\cdots$ 123132$\cdots$132 for odd $n$, and 123$\cdots$231213$\cdots$132312$\cdots$123 for even $n$, respectively. From the above observations, it is shown that $\varphi^{n+1} (2)$ does not contain any square with the length either 2 or 4.  

\underline{\textit{Case 2}:\ \ $\ell\equiv 0 \ (\textrm{mod}\ 3)$}

In this case, if $\varphi^{n+1} (2)$ contains a square with the length $2\ell$, $\varphi^n (2)$ contains a square with the length $2\ell/3$, according to Lemma \ref{central}. This contradicts the assumption that $\varphi^n (2)$ is square-free. 

\underline{\textit{Case 3}:\ \ $\ell\ge 4$ and $\ell\not\equiv 0 \ (\textrm{mod}\ 3)$}

In this case, the proof is rather complicated than those in the other cases. It is necessary to further divide the cases, according to the values of $\ell$ and $p$. Here, we omit a superscript $n+1$ from $a_i^{n+1}$ for simplicity: $\varphi^{n+1}(2)=\{a_i\}_{i=0}^{3^{n+1}-1}$.
 
For example, we consider the case that $\ell=3m+1$ and $p=3q$ where $m$ and $q$ are certain integers within an appropriate range. Assume that $\varphi^{n+1} (2)$ contains a square such that $a_k=a_{k+3m+1}$ for $k=3q,3q+1,\cdots,3(q+m)$. Note a couple of ternarys $a_{3q}a_{3q+1}a_{3q+2}$ and $a_{3(q+m)}a_{3(q+m)+1}a_{3(q+m)+2}$. One can deduce that $a_{3q+2}=a_{3(q+m)}$ because $a_{3q}=a_{3(q+m)+1}$ and $a_{3q+1}=a_{3(q+m)+2}$ from the assumption, and Lemma \ref{ternary} guarantees that $a_{3q+2} \ (a_{3(q+m)})$ is different from $a_{3q}$ and $a_{3q+1}$ ($a_{3(q+m)+1}$ and $a_{3(q+m)+2}$). The similar deductions for each couple of ternarys lead to 

\[
\hspace{-5mm}
a_{3q+2}=a_{3q+5}=\cdots=a_{3(q+m)-1}=a_{3(q+m)}=a_{3(q+m+1)}=\cdots=a_{3(q+2m)},
\]

and hence $\varphi^{n+1} (2)$ contains a square with the length 2: $a_{3(q+m)-1}=a_{3(q+m)}$.
 
For another example, we consider the case that $\ell=3m+2$ and $p=3q+1$, and assume that $\varphi^{n+1} (2)$ contains a square such that $a_k=a_{k+3m+2}$ for $k=3q+1,3q+2,\cdots,3(q+m)+2$. $a_{3q}=a_{3(q+m+1)+2}$ is deduced noting a couple of ternarys $a_{3q}a_{3q+1}a_{3q+2}$ and $a_{3(q+m+1)}a_{3(q+m+1)+1}a_{3(q+m+1)+2}$. It follows that

\[
\hspace{-5mm}
a_{3q}=a_{3q+3}=\cdots=a_{3(q+m)}=a_{3(q+m+1)+2}=a_{3(q+m+2)+2}=\cdots=a_{3(q+2m)+2}.
\]

Note a couple of consecutive ternarys $a_{3(q+m)}a_{3(q+m)+1}a_{3(q+m)+2}$ and $a_{3(q+m+1)}$ $a_{3(q+m+1)+1}a_{3(q+m+1)+2}$ in turn.
Since $\varphi^n(2)$ is square-free, Lemma \ref{central} allows one to deduce $a_{3(q+m)+1}\neq a_{3(q+m+1)+1}$. Combine this fact and $a_{3(q+m)}=a_{3(q+m+1)+2}$, and then $a_{3(q+m)+1}=a_{3(q+m+1)}, a_{3(q+m)+2}=a_{3(q+m+1)+1}$ from Lemma \ref{ternary}. Therefore, $\varphi^{n+1} (2)$ contains a square with the length 4.

Under modulo 3, $\ell$ and $p$ are divided into two and three cases, respectively. There exist six cases in total. We can prove that, if $\varphi^{n+1} (2)$ contains a square, it contains another square with the length 2 or 4 at the joint of the square, for other four cases as well as for the above two cases. To summarize, Case 3 can be reduced to Case 1, by utilizing Lemmas \ref{ternary} and \ref{central}.

\vspace{4mm}

From the above considerations for three cases, we can conclude that $\varphi^{n+1}(2)$ does not contain a square with any length $2\ell$, and thus Theorem \ref{square-free} has been established by a mathematical induction. \qed
   
\end{pf}

Note that $\varphi^\infty (2)$ does not yet converge toward a definite sequence. We should alter the numbering of elements of $\varphi^n (2)$, in order to derive an infinite sequence. Taking the symmetry of the ternary into account, it is required to employ a full-infinite sequence, instead of a semi-infinite sequence. When we set the numbering of elements of $\varphi^n (2)$ in such a way that $\varphi^n (2)\equiv \{a_i\}_{i=-(3^n-1)/2}^{(3^n-1)/2}$, the sequence $\varphi^n (2)$ is included in the middle of $\varphi^{n+1} (2)$, and then the (full-) infinite square-free sequence $\varphi^\infty (2)\equiv \{a_i\}_{i=-\infty}^{\infty}$ can be defined. $\varphi^\infty (2)$ is quite similar to the Arshon sequence \cite{Arshon}, but essentially different, because the Arshon sequence is semi-infinite.

\section{Direct definition of a ternary infinite square-free sequence}
\label{}

 In this section, we redefine the square-free sequence derived in section \ref{Const}, by a direct definition. The key to a direct definition is to represent an integer $i$ not by an ordinary ternary representation but by a \textit{balanced} ternary representation. The latter uses the digits $-1,0,+1$ whereas the former uses the digits $0,1,2$.
 
 Let the balanced ternary representation of an integer $i$ ($-\infty<i<\infty$) be
\begin{equation}
i=\sum_{n=0}^{\nu} u_n 3^n \equiv [u_{\nu}\cdots u_1u_0]_3
\end{equation}
where $u_n \in \Sigma =\{ -1,0,+1 \}$ and $u_{\nu} \neq 0$. Note that this representation of $i$ is uniquely determined. 

Here, permutations $\pi_{-1},\pi_0,\pi_{+1}$ over $\{-1,0,+1 \}$ are defined in the following manner, respectively:

\begin{itemize}
	\item $\pi_{-1}$ exchanges $-1$ and 0.
	\item $\pi_0$ is the identical permutation.
	\item $\pi_{+1}$ exchanges 0 and $+1$.
\end{itemize}
In terms of $\pi_{-1},\pi_0,\pi_{+1}$, we define the infinite sequence $\{b_i\}_{i=-\infty}^{\infty} \in \{-1,0,+1 \}^*$ such that
\begin{equation}
\label{BF_ternary}
b_i=\pi_{u_{\nu}}\circ \cdots \circ \pi_{u_1}\circ \pi_{u_0}(0)
\end{equation}
with $i=[u_{\nu}\cdots u_1u_0]_3$. For example, $b_8=\pi_{+1}(\pi_0(\pi_{-1}(0)))=-1$ and $b_{-17}=\pi_{-1}(\pi_{+1}(\pi_0(\pi_{+1}(0))))=-1$ because $8=3^2-1$ and $-17=-3^3+3^2+1$, respectively.

Finally, we redefine $\sigma$ and $\rho$ on $\{-1, 0, +1\}^*$. It follows that $\varphi^n (0)$ denotes the sequence obtained from $\varphi^n (2)=\{a_i\}_{i=-(3^n-1)/2}^{(3^n-1)/2}$ by replacing 1,2 and 3 with $-1$, 0 and $+1$, respectively. 

Based on the above notations, a direct definition of a ternary infinite square-free sequence is derived:
\begin{thm}
\label{direct_definition}
$\{b_i\}_{i=-(3^n-1)/2}^{(3^n-1)/2}$ is equivalent to $\varphi^n (0)$ for arbitrary integer $n$. Therefore, the ternary infinite sequence $\{b_i\}_{i=-\infty}^{\infty}$ is square-free according to Theorem \ref{square-free}.
\end{thm}

\def\Elproofname{PROOF.}
\begin{pf}
The proof proceeds via a mathematical induction. For $n=1$, it is obvious that $b_{-1}b_0b_{+1}=\varphi(0)$. Assume that $\{b_i\}_{i=-(3^n-1)/2}^{(3^n-1)/2} = \varphi^n (0)$. For $(3^n-1)/2< i\le (3^{n+1}-1)/2$, $i=3^n+i'$ where $|i'|\le (3^n-1)/2=3^{n-1}+3^{n-2}+\cdots +1$, namely, the highest digit of $i$ in the balanced ternary representation is 1. Then, we obtain $\{b_i\}_{i=(3^n-1)/2+1}^{(3^{n+1}-1)/2}=\{\pi_{+1}(b_{i'})\}_{i'=-(3^n-1)/2}^{(3^n-1)/2}=\rho(\varphi^n (0))$. Similarly, we obtain $\{b_i\}_{i=-(3^{n+1}-1)/2}^{-(3^n-1)/2-1}=\sigma(\varphi^n (0))$. It follows that

\begin{equation}
\{b_i\}_{i=-(3^{n+1}-1)/2}^{(3^{n+1}-1)/2}=  \sigma(\varphi^n (0))\ \varphi^n (0)\ \rho(\varphi^n (0))=\varphi^{n+1} (0).
\end{equation}
Therefore, the proof of Theorem \ref{direct_definition} has been completed. \qed

\end{pf}

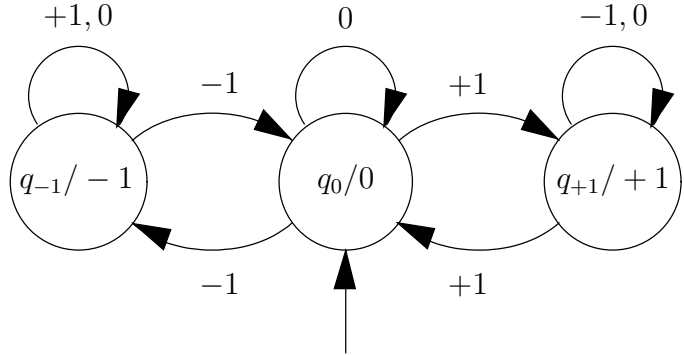
\begin{figure}[htbp]
\vspace{0.2cm}
\begin{center}
\unitlength 0.1in
\begin{picture}(36.60,18.42)(0.55,-22.07)
%
\special{pn 8}%
\special{ar 1960 1310 349 358  0.0000000 6.2831853}%
%
\special{pn 8}%
\special{sh 1.00}%
\special{pa 3068 1085}%
\special{pa 2880 1036}%
\special{pa 2927 934}%
\special{pa 3068 1085}%
\special{pa 3068 1085}%
\special{pa 3068 1085}%
\special{fp}%
%
\special{pn 8}%
\special{pa 1960 1668}%
\special{pa 1960 2207}%
\special{fp}%
%
\special{pn 8}%
\special{sh 1.00}%
\special{pa 850 1530}%
\special{pa 1044 1561}%
\special{pa 1006 1666}%
\special{pa 850 1530}%
\special{pa 850 1530}%
\special{pa 850 1530}%
\special{fp}%
%
\special{pn 8}%
\special{sh 1.00}%
\special{pa 2231 1525}%
\special{pa 2425 1556}%
\special{pa 2387 1661}%
\special{pa 2231 1525}%
\special{pa 2231 1525}%
\special{pa 2231 1525}%
\special{fp}%
%
\special{pn 8}%
\special{sh 1.00}%
\special{pa 762 1022}%
\special{pa 773 822}%
\special{pa 879 851}%
\special{pa 762 1022}%
\special{pa 762 1022}%
\special{pa 762 1022}%
\special{fp}%
%
\special{pn 8}%
\special{sh 1.00}%
\special{pa 2160 1022}%
\special{pa 2171 822}%
\special{pa 2277 851}%
\special{pa 2160 1022}%
\special{pa 2160 1022}%
\special{pa 2160 1022}%
\special{fp}%
%
\special{pn 8}%
\special{sh 1.00}%
\special{pa 3558 1022}%
\special{pa 3569 822}%
\special{pa 3675 851}%
\special{pa 3558 1022}%
\special{pa 3558 1022}%
\special{pa 3558 1022}%
\special{fp}%
%
\special{pn 8}%
\special{ar 562 861 262 260  2.5427048 6.2831853}%
\special{ar 562 861 262 260  0.0000000 0.6207307}%
%
\special{pn 8}%
\special{sh 1.00}%
\special{pa 1671 1085}%
\special{pa 1482 1036}%
\special{pa 1530 934}%
\special{pa 1671 1085}%
\special{pa 1671 1085}%
\special{pa 1671 1085}%
\special{fp}%
%
\special{pn 8}%
\special{ar 1960 861 262 260  2.5427048 6.2831853}%
\special{ar 1960 861 262 260  0.0000000 0.6207307}%
%
\special{pn 8}%
\special{ar 3357 861 263 260  2.5397638 6.2831853}%
\special{ar 3357 861 263 260  0.0000000 0.6212092}%
%
\special{pn 8}%
\special{pa 1680 1525}%
\special{pa 1656 1546}%
\special{pa 1629 1564}%
\special{pa 1603 1581}%
\special{pa 1575 1597}%
\special{pa 1546 1611}%
\special{pa 1516 1623}%
\special{pa 1486 1634}%
\special{pa 1455 1643}%
\special{pa 1424 1651}%
\special{pa 1393 1657}%
\special{pa 1361 1662}%
\special{pa 1329 1664}%
\special{pa 1298 1668}%
\special{pa 1266 1668}%
\special{pa 1234 1668}%
\special{pa 1202 1666}%
\special{pa 1170 1662}%
\special{pa 1138 1659}%
\special{pa 1107 1653}%
\special{pa 1076 1645}%
\special{pa 1045 1637}%
\special{pa 1014 1627}%
\special{pa 985 1614}%
\special{pa 955 1602}%
\special{pa 928 1586}%
\special{pa 900 1570}%
\special{pa 873 1552}%
\special{pa 849 1532}%
\special{pa 842 1525}%
\special{sp}%
%
\special{pn 8}%
\special{pa 3077 1525}%
\special{pa 3053 1546}%
\special{pa 3027 1564}%
\special{pa 3000 1581}%
\special{pa 2972 1597}%
\special{pa 2943 1611}%
\special{pa 2913 1623}%
\special{pa 2883 1634}%
\special{pa 2853 1643}%
\special{pa 2822 1651}%
\special{pa 2790 1657}%
\special{pa 2759 1662}%
\special{pa 2727 1665}%
\special{pa 2695 1668}%
\special{pa 2663 1668}%
\special{pa 2631 1668}%
\special{pa 2599 1666}%
\special{pa 2567 1663}%
\special{pa 2535 1659}%
\special{pa 2504 1653}%
\special{pa 2473 1645}%
\special{pa 2442 1637}%
\special{pa 2412 1626}%
\special{pa 2382 1614}%
\special{pa 2353 1601}%
\special{pa 2325 1586}%
\special{pa 2298 1569}%
\special{pa 2270 1552}%
\special{pa 2246 1531}%
\special{pa 2239 1525}%
\special{sp}%
%
\special{pn 8}%
\special{ar 1261 1310 523 359  3.7839483 5.6408297}%
%
\special{pn 8}%
\special{ar 2658 1310 524 359  3.7855521 5.6399131}%
%
\special{pn 8}%
\special{ar 562 1310 358 358  0.0000000 6.2831853}%
%
\special{pn 8}%
\special{ar 3357 1310 358 358  0.0000000 6.2831853}%
\put(19.6000,-13.0000){\makebox(0,0){$q_0/0$}}%
\put(5.5000,-13.0000){\makebox(0,0){$q_{-1}/-1$}}%
\put(33.6000,-13.0000){\makebox(0,0){$q_{+1}/+1$}}%
\put(26.0000,-8.0000){\makebox(0,0){$+1$}}%
\put(26.0000,-18.5000){\makebox(0,0){$+1$}}%
\put(13.0000,-8.0000){\makebox(0,0){$-1$}}%
\put(13.0000,-18.5000){\makebox(0,0){$-1$}}%
\put(19.6000,-4.5000){\makebox(0,0){$0$}}%
\put(33.6000,-4.5000){\makebox(0,0){$-1,0$}}%
\put(5.6000,-4.5000){\makebox(0,0){$+1,0$}}%
%
\special{pn 8}%
\special{sh 1.00}%
\special{pa 1960 1670}%
\special{pa 1900 1870}%
\special{pa 1960 1870}%
\special{pa 2010 1870}%
\special{pa 2010 1870}%
\special{pa 1960 1670}%
\special{fp}%
\end{picture}%
\end{center}
\caption{DFAO generating the sequence $\{b_i\}_{i=-\infty}^\infty$}
\label{fig:DFAO}
\end{figure}

The formula (\ref{BF_ternary}) can be regarded as the definition of a DFAO which takes as input the balanced ternary representation of an integer $i$, read in order from low to high digits. Indeed, the three function $\pi_a$ denote the state transition function among the three state $q_a/a$ ($a=-1,0,+1$), where, by convention, the state labeled $q_a/a$ indicates that the output associated with the state $q_a$ is $a$ and the initial state is $q_0/0$. The state transition diagram for this DFAO is drawn as a directed graph in Figure \ref{fig:DFAO}. Therefore, the sequence $\{b_i\}_{i=-\infty}^\infty$ is automatic as well as the Arshon sequence.

The two equivalent ways of the definition of the ternary infinite square-free sequence, $\varphi^\infty(0)$ and $\{b_i\}_{i=-\infty}^{\infty}$, are almost parallel with those of the Thue-Morse sequence. $\varphi$ is to $\varphi^\infty (0)$ what $\psi$ is to the Thue-Morse sequence $\psi^\infty(0)$. Furthermore, the balanced ternary representation is to the function $b_i$ (\ref{BF_ternary}) what the binary representation is to the function $s_2 (i)$. In this sense, our sequence is said to be the counterpart of the Thue-Morse sequence, i.e., a natural extension from binary to ternary.

\section{Conclusion and discussion}
\label{Concl}

We have proposed a novel construction of the ternary infinite square-free sequence, generated by the automaton operation with the balanced ternary representation of an integer. The sequence presented in this paper may be favorable for various applications, because of its easiness to generate. The nature of \textit{square-free} is intriguing and useful itself. We hope that the properties of our sequence are further clarified and utilized in various fields such as information science, random system on physics and quasiperiodic tiling. 

\section*{Acknowledgement}

The author (TK) acknowledges Prof. M. Wadati for the critical reading of this manuscript. He is also grateful to Prof. S. Yamada for pointing out the possibility of the definition such as (\ref{BF_ternary}), which helps strikingly to raise up the value of this work. 


\end{document}